\documentclass{elsarticle}

\usepackage{xcolor}
\usepackage{comment}

\usepackage{subcaption}
\usepackage{graphicx}

\usepackage{booktabs}
\usepackage{url}

\usepackage[utf8]{inputenc}
\usepackage{booktabs}
\usepackage{dcolumn,tabularx,ragged2e}
\usepackage{siunitx}
\usepackage{tablefootnote}

\usepackage{multirow,tabularx}
\usepackage{longtable}

\usepackage[T1]{fontenc}
\usepackage{textcomp}
\usepackage{listings}

\newif\ifdraft
\drafttrue

\usepackage{xcolor}

\journal{Elsevier Journal of Systems and Software}

\begin{document}

\begin{frontmatter}

\title{Evolution of repositories and privacy laws: commit activities in the GDPR and CCPA era}

\author[1]{Georgia M. Kapitsaki\corref{cor1}}
\ead{gkapi@ucy.ac.cy}
\author[1]{Maria Papoutsoglou}
\ead{mpapoutsogloy@gmail.com}

\cortext[cor1]{Corresponding author}

\address[1]{University of Cyprus, 1, University Avenue, Aglantzia, 2109, Cyprus}

\begin{abstract}
Free and open source software has gained a lot of momentum in the industry and the research community. The latest advances in privacy legislation, including the EU General Data Protection Regulation (GDPR) and the California Consumer Privacy Act (CCPA), have forced the community to pay special attention to users' data privacy. The main aim of this work is to examine software repositories that are acting on privacy laws. We have collected commit data from GitHub repositories in order to understand indications on main data privacy laws (GDPR, CCPA, CPRA, UK DPA) in the last years. Via an automated process, we analyzed 37,213 commits from 12,391 repositories since 2016, whereas 594 commits from the 70 most popular repositories of the dataset were manually analyzed. We observe that most commits were performed on the year the law came into effect and privacy relevant terms appear in the commit messages, whereas reference to specific data privacy user rights is scarce. The study showed that more educational activities on data privacy user rights are needed, as well as tools for privacy recommendations, whereas verifying actual compliance via source code execution is a useful direction for software engineering researchers. 
\end{abstract}


\end{frontmatter}

\maketitle

\section{Introduction}

Privacy relevant laws that have come into effect in the last 6 years include predominantly the EU General Data Protection Regulation (GDPR)~\cite{voigt2017eu} and the California Consumer Privacy Act of 2018 (CCPA)~\cite{pardau2018california}. California Privacy Rights Act (CPRA)~\cite{determann2020california} is a recent amendment to CCPA, whereas other countries are also following the rationale of introducing similar privacy laws, such as the Data Protection Act of UK (UK DPA)~\cite{uk_dpa_2018} that is the UK’s implementation of the GDPR, the Brazilian General Personal Data Protection Law~\cite{brazil_lgpd} and the New Zealand's Privacy Act~\cite{dong2020privacy}. 

A vast amount of software repositories are available on GitHub and they are being reused by developers worldwide. Since software may also contain mechanisms for users' data collection, the community is obliged to adhere to relevant legislation in order to protect users' privacy~\cite{li2019impact}. Overall, a positive impact of the privacy laws on practitioners' behaviours and organizations' cultures has been reported~\cite{iwaya2023privacy}. Previous works have examined developers' concerns on privacy in specific domains, e.g. analyzing Reddit discussions on mobile application development~\cite{li2021developers,parsons2023understanding}, Q\&A sites on privacy~\cite{tahaei2020understanding} or pull requests~\cite{franke2024first}, but have not investigated the repositories' evolution when it comes to commits. However, the question of how privacy laws have affected software repositories so that they act on updating their source code is a useful area of research to allow us to understand whether repositories have performed commits for data privacy laws, and how long it took to make appropriate changes, so that tendencies can be found and areas of improvement can be identified.

To guide our research in an attempt to answer the above basic question, we structure our work around the following research questions (RQs):
\begin{itemize}
    \item \textbf{RQ1}. \emph{Which} mentions to main data privacy laws \emph{do commits make and how has this evolved over time?} We have analyzed the number of commits and the changed Lines of Code (LOC) in commits that make reference to main data privacy laws, observing the year the change was made, as well as how long it took to complete these relevant commits. GDPR, CCPA, CPRA and UK DPA are the laws considered in this and the next RQs.% We examined when was the first and the last relevant commit in repositories.
    \item \textbf{RQ2}. \emph{Which type of repositories and which programming languages} are more active in relevant development activities on data privacy legislation? In this RQ, our intent was to find how many commits are performed in repositories and how many LOC are usually changed. The effect of the owner type (User or Organization) and of the main programming language of the repository on the commits volume is also examined.
    \item \textbf{RQ3}. \emph{What are the main terms appearing together with data privacy laws in commits on GitHub?} As there are relevant privacy terms that may be indicated in the commits, we examined which are the most frequent terms appearing in the commit message per examined law and the presence of specific user rights from the legislation. 
    \item \textbf{RQ4}. \emph{Which is the main purpose of a commit that addresses privacy legislation?} The commit may refer to code change or to text change (e.g. updates to privacy policy text). In order to complement the automated analysis that was used in the previous three RQs, the aim of the RQ is to examine some commit messages in more detail. The RQ was answered by manually analyzing the commit messages and relevant code changes from a small subset of the dataset (594 commits from the 70 most popular repositories in the dataset). 
\end{itemize}

We collected commits using keywords from recent privacy laws: GDPR, CCPA, CPRA, UK DPA. Via a mainly automated and partially manual process, we analyzed 36,807 commits from 12,391 repositories from 2016 till 2023. To the best of our knowledge, no study on commit activities for privacy legislation on GitHub commits exists and no related study involving commits has utilized such a large number of commits coming from a large number of repositories. The contribution of our work is summarized in the following: i) creation of a huge labeled corpus of commits related to data privacy laws, ii) analysis of references to data privacy laws in commits on GitHub using commits volume and changed LOC, iii) investigation of the characteristics of the relevant commits and repositories (e.g. main terms). An initial version of this line of work has been published with a preliminary analysis on the indication of GDPR in GitHub commits~\cite{kapitsaki2024gdpr}.

The remainder of the text is structured as follows. Section 2 presents relevant works in the area. The methodological process is introduced in section 3. Results are presented in section 4 and are further discussed together with implications in section 5. Threats to validity are examined in section 6, and finally, section 7 concludes the work.

\section{Background and related work}

\subsection{Privacy laws}

The landscape of privacy legislation has changed primarily after the introduction of GDPR that came into effect in May 25th, 2018 and superseded the Data Protection Directive 95/46/EC~\cite{voigt2017eu}. UK's DPA is the UK version of GDPR that complements GDPR to suit the needs of UK (with the same effect date as GDPR). CCPA followed in the United States coming into effect on January 1st, 2020, whereas CPRA adds to the existing provisions of CCPA (effect date: January 1st, 2023)~\cite{determann2020california}. Many other countries have followed the GDPR paradigm introducing recently data privacy laws. According to the United Nation's Conference on Trade and Development\footnote{\url{https://unctad.org/page/data-protection-and-privacy-legislation-worldwide}}, 71\% of countries worldwide have data protection and privacy legislation in place with a total of 241 laws including older and newer laws within one country.

Software needs to comply with the above laws, provided that it is active in the geographical areas the law applies. For instance, the INFORM e-learning platform was implemented considering GDPR requirements~\cite{vanezi2019gdpr}. One main area on data collection is that users need to provide their consent on the collecting, usage and processing of their personal data and they are usually informed about these aspects via the privacy policy of the service or application. User rights are fundamental, as they provide users control over data activities. GDPR indicates 8 user rights (also present in UK's Data Protection Act): the right to information, the right of access, the right to rectification, the right to erasure, the right to restriction of processing, the right to data portability, the right to object, and the right to avoid automated decision-making. CCPA has introduced five main privacy rights for consumers: the right to know, the right to delete, the right to opt-out (of sale), the right to disclosure, and the right to non-discrimination, and CPRA four additional rights: the right to correct (inaccurate personal information), the right to opt-out of automated decision making, the right to data portability and the right to limit use and disclosure of sensitive personal information. UK DPA shares the same rights with GDPR.

\subsection{Commit/issue analysis}

How Free and Open Source Software is used by large organizations was examined using 1,314 repositories from GitHub~\cite{chelkowski2021free}. Specific metrics were used for this, including frequency of commits, Lines of Code and comments in source code. The relation of commits' sentiment in relation to software bugs was investigated, with a main conclusion that commits related to bugs (introducing, preceding or fixing bugs) are more negative than other types of commits~\cite{huq2020developer}. Issue comments on GitHub were analyzed by Khalajzadeh et al. with the aim of identifying human-centric issues and a wide range was encountered, including \emph{Privacy \& Security}~\cite{khalajzadeh2022diverse}. Other works on commits analysis include the Anomalicious tool that assists in detecting potentially malicious commits~\cite{gonzalez2021anomalicious}, works that detect unusual commits~\cite{goyal2018identifying}, as well as works that perform commit classification~\cite{casalnuovo2017gitcproc,gharbi2019classification,macho2016predicting} or works that examine other properties (e.g. size of commits, software design degradation)~\cite{hattori2008nature,ferreira2022characterizing,oliva2013can}.  
Labeling issues as questions, bugs, enhancements has been examined using BERT (Bidirectional Encoder Representations from Transformers)~\cite{siddiq2022bert}. Issues were also studied in earlier works to study the overall adoption of issue trackers, the relevant categories, how they are used by the project's community and how they relate with the project's success~\cite{bissyande2013got}.

\subsection{Data privacy in Web applications and Open Source Software}
 
Topic modelling was applied to 1,733 privacy-related questions on Stack Overflow and a random sample of 315 questions was then qualitatively analyzed by Tahaei et al.~\cite{tahaei2020understanding}. Questions collected included the word `privacy' either in their title or as a tag. Laws and regulations, such as GDPR, were included among the themes in a thematic analysis performed in the qualitative examination to identify the drivers that made the user post a specific question. Regulations outside EU were not found in that sample but in the whole dataset references to other regulations, such as USA’s
Health Insurance Portability and Accountability Act (HIPAA), were found. A similar work that employed topic modelling considering also Information Security and Software Engineering Stack Exchange sites (apart from Stack Overflow) was performed by Diepenbrock et al.~\cite{diepenbrock2023analysis}. Among the topics, the \emph{Legal} topic covers discussions on compliance with laws, such as GDPR and CCPA.
 
Developers' discussions on Reddit were examined using a qualitative analysis of a sample of 207 threads mentioning different forms of personal data from the r/androiddev forum on Reddit~\cite{li2021developers}. The authors relied on the legislation including GDPR and CCPA to extract relevant terms for personal data. Another work extended the previous analysis on Reddit discussions and used word frequency, topic clustering and classification to analyse 437,317 threads from r/webdev, r/androiddev, and r/iOSProgramming~\cite{parsons2023understanding}. Concerning GDPR and CCPA, it was found that there is a significant change in topics and terms due to GDPR but to a lesser extent due to CCPA. 

In order to see whether websites are complying with the minimum requirements of CCPA, providing a link to hyperlink on their homepage with the text ``Do Not Sell
My Personal Information" (DNSMPI) (or right to opt-out of sale), a corpus of web documents was examined~\cite{van2022setting}. Developers of mobile applications directed to children were asked about the privacy compliance processes they follow, including reference to the developers' perspectives on the requirements of Children's Online Privacy Protection Act (COPPA), GDPR and CCPA~\cite{alomar2022developers}. It was found that developers put a lot of trust in the enforcement performed in the application markets and as a result there is a need for more usable compliance-checking and auditing tools. Online, various approaches are investigating privacy policies automating their analysis~\cite{harkous2018polisis} or examining compliance to legislation, such as GDPR~\cite{vanezi2021complicy,DBLP:conf/webist/KailiK22}.

Previous works have also identified the need to investigate how software developers address privacy regulations~\cite{mazeli2022framework}. Privacy-relevant requirements for software projects were exported in a taxonomy from GDPR, the ISO/IEC 29100 privacy framework, the Thailand Personal Data Protection Act and the Asia-Pacific Economic Cooperation (APEC) privacy framework~\cite{sangaroonsilp2023taxonomy}. The issue reports of Chrome and Moodle were also classified in the taxonomy, and some differences between privacy and non-privacy issues were found. A framework was described to examine in the future which issues developers discuss in relevance to privacy legislation~\cite{henning2023understanding}. The authors identified the relevant gap in the literature and intend to work on the reporting of issues related to personal data and data protection. In their work, they describe their analysis plan that focuses mainly on issue types and reporter types analysis. Pull requests were analyzed along with the results of a survey with developers to understand the effect of GDPR on open source software development~\cite{franke2024first,franke2024exploratory}. Main results include that there is more development activity with GDPR-related pull requests in terms of commits, additions, deletions, and files changed, as well as review activity, while no variations were found in the sentiment of pull requests over time.\\

\textbf{Relation to previous works.} No previous work has examined whether GitHub repositories are acting on accommodating the needs of recent and popular data privacy laws via commits analysis, whereas only one work has focused on pull requests~\cite{franke2024first,franke2024exploratory}. An important advantage of the current work is that it relies on all available commits on GitHub and is not restricted to specific repositories, programming languages or ecosystems.

\section{Methodology steps}

\begin{figure*}[!t]
\centering
\includegraphics[scale=0.5]{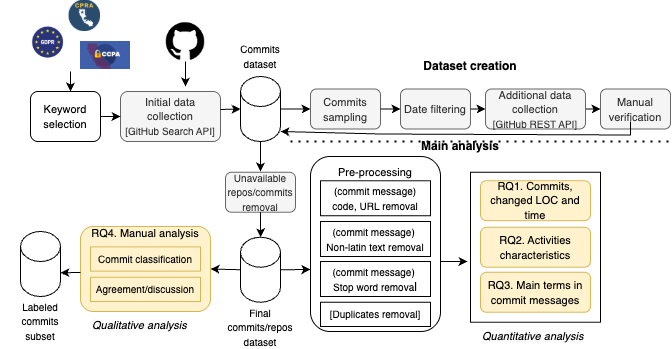}
\caption{Methodological steps.}
\label{fig:methodological-process} 
\end{figure*}

The phases used in the work are described in the next parts. The whole procedure is depicted in~\figurename~\ref{fig:methodological-process}. We have used commits as main source for analysis, instead of e.g. issues, for the following reasons. First, some repositories may be using external issue tracking systems (e.g. Jira) or may not be using any issue tracking system, so if we relied on issues we would not have been able to collect information on such repositories. A previous work has reported that issues are almost exclusively reported in large projects with big development teams~\cite{bissyande2013got}. Secondly, issue discussions have a more complex structure than commits, and performing the search based on privacy legislation may have included irrelevant content in the dataset or may have excluded relevant data (e.g. there is a mention of GDPR in comments of an issue\footnote{\url{https://github.com/jedisct1/libsodium.js/issues/24}} but the issue is not relevant to privacy legislation). Analyzing issues is nevertheless, an important area of work on data privacy legislation~\cite{henning2023understanding}.

\subsection{Keyword selection from legislation}

In order to choose the keywords of interest to search for in projects' commits, we relied on the experience of one of the authors on previous works on Privacy Enhancing Technologies and GDPR compliance in software systems in the last 12 years~\cite{kounoudes2020mapping,vanezi2024s} and considered the relevant privacy laws~\cite{voigt2017eu,pardau2018california,determann2020california,uk_dpa_2018}. We devised the following set of keywords that needed to be present in the commit: \emph{General Data Protection Regulation} or its abbreviation \emph{GDPR}, \emph{California Consumer Privacy Act} or its abbreviation \emph{CCPA}, \emph{California Privacy Rights Act} or its abbreviation \emph{CPRA}, and finally \emph{Data Protection Act}. We chose the above laws for the following reasons. They are all recent and popular with software needing to comply when offered within the laws' jurisdiction~\cite{horak2019gdpr}. Second, they have gained popularity in research works: a search on GDPR on Google Scholar since 2022 (and till March 2025) returns approximately 20,900 publications, on CCPA 49,400 and on CPRA 3,810 (the last two abbreviations may though be used also for other purposes). Finally, as aforementioned one of the authors has research experience with one of the laws (GDPR). 

We did not consider laws that are less popular on a global level, e.g. the Brazilian General Data Protection Act effective since February 2020 does not appear in any commit. We also did not include the abbreviation of \emph{Data Protection Act}, DPA, or more generic privacy relevant terminology, such as \emph{data privacy}, as the amount of search results is very large (552,000 commits returned for \emph{DPA} and 17,400 commits returned for \emph{data privacy} via GitHub search in March 2025) and they may not be linked to privacy laws relevant changes that is our focus. DPA is also used for other purposes, and also as abbreviation e.g. for Digital Process Automation, as a check of some commits shows:
\begin{itemize}
    \item Fluepke/luca-web-clone: ``\emph{new checkbox for data processing agreement (DPA) in registration}" 
 [SHA:2f878ef9e624224722aa073ee71cb8703f6728f1]
    \item iqrfsdk/clibdpa: ``\emph{Update DPA.h for 410}" \hfill \break
 [SHA:51f9b9ddb062d7c6857135c911ee9a3ccef84245]
    \item dfath/simbapha: ``\emph{
basic volume dpa}" \hfill \break[SHA:9cacef1a8f32b6e68ec2a6c6adefdb936d1a558a]
\end{itemize}
We also did not focus on security discussions on GitHub, as performed on previous works and have thus, excluded other keywords such as \emph{encryption}, \emph{authentication}, \emph{authorization}~\cite{zhou2021spi,zahedi2018empirical,buhlmann2022developers}.

\subsection{Dataset collection and preliminary analysis}  

\textbf{Initial data collection.} Table~\ref{tab:dataset-size} shows the initial size of the dataset that we collected per keyword in April, 2023 (columns: \emph{collected} and \emph{keyword} sum, no duplicates, after summing the results for each law keyword, e.g. \emph{General Data Protection Regulation} and \emph{GDPR}). Data were collected using the GitHub Search API,\footnote{\url{https://docs.github.com/en/rest}} with a search within the text (and comments) of commits. It was chosen as the most suitable option, as we were interested in collecting up to date data~\cite{mombach2018github}. Date intervals were also indicated in the request to ensure that all commits for each keyword would be collected. Since GitHub Search API uses a limit of 1,000 results per search query, we used date and time intervals and called the API numerous times to ensure we would collect all relevant data.

\begin{table}
\centering
  \caption{Dataset size per keyword.}
  \label{tab:dataset-size}
\resizebox{\columnwidth}{!}{%
  \begin{tabular}{lrrrrr}
    \toprule
    \textbf{Keyword} & \multicolumn{5}{c}{\textbf{\# commits}}\\
    & \textbf{collected} & \textbf{keyword} & \textbf{date} & \textbf{removed} & \textbf{final}\\
    &&&\textbf{sum, no} &\textbf{filter} & \textbf{unavailable}\\    
    & &\textbf{duplicates}\\
    \hline 
    GDPR & 72,729 & 35,910 & 35,900 & 35,254 & 35,254\\
    General Data Protection Regulation & 509 &&&&\\
    CCPA & 3,011 & 2,980 & 2,957 & 1,894 & 1,863\\
    California Consumer Privacy Act & 56 \\
    CPRA & 212 & 212 & 69 & 69 & 35\\
    California Privacy Rights Act & 0 \\
    Data Protection Act & 100 & 86 & 67 & 67 & 61\\
    \hline
    \textbf{TOTAL \# commits} &&&&& \textbf{37,213}         \\
    &&\multicolumn{4}{r}{\textbf{(36,807} no duplicates)}\\
\hline
\textbf{TOTAL \# repositories} &&&&& \textbf{12,391}\\
\hline
\end{tabular}%
}
\end{table}

\textbf{Duplicates removal.} We removed duplicates using the commit SHA, as we observed that different repositories may have the same commit that has been kept most probably because of forked repositories that have later evolved. We first ordered the commits based on the repository stars number as a proxy of popularity and then removed the duplicates, so less popular repositories with the same commits were filtered out. Stars show whether a user likes the repository or wants to show her appreciation and are a popularity indicator that has also been used in previous works~\cite{borges2016understanding,borges2018s}.

\textbf{Date filtering.} By inspecting a small number of commit messages, 100 random commits irrespective of the keyword (one of the authors performed this task), we found irrelevant cases, e.g. for CCPA there was reference to a clock tree and irrelevant mentions including ccpa in the name,\footnote{\url{https://github.com/dlax/sido-mmc/commit/8bd2993ebc3996e7de46a1efac0e25a0c4b741d6}} CPRA was used to refer to stations.\footnote{\url{https://github.com/CPRA-MP/ICM_Hydro/commit/49219b2b16c6570e3bd59ac69709a812dc4d210d}} This manual inspection was performed by reading the commit message and, when necessary, by visiting additionally the commit page and examining the changed source code. In order to make sure not to include such cases, we filtered out references to a specific keyword that were made before the signing of the respective law (not its effect date that is later than the dates indicated next): before April 14th, 2016 for GDPR, June 28th, 2018 for CCPA, November 3rd, 2020 for CPRA, May 23rd, 2018 for the UK Data Protection Act. The excluded commits were then verified by another author and no cases that were falsely excluded were encountered apart from one commit indicating CPRA with a date very close to the signing of the law (Oct. 29th, 2020).\footnote{\url{https://github.com/aleecia/CCPABrowserTool/commit/fc795a994d6284633393ca940796ce82fd69f702}} The size of the dataset after the date filtering is shown in the \emph{date filter} column of Table~\ref{tab:dataset-size}.

\textbf{Additional data collection and unavailable data removal.} Since the response of the GitHub Search API did not include all fields needed for our analysis, we collected additional data on repository and on commit level. The data used from each API call type and kept for the subsequent analysis are listed in Table~\ref{tab:data-used}. They are coming from 3 different call types: GitHub Search API, GitHub REST API repository information resource, GitHub REST API commit information resource. However, when we proceeded with the repository information collection via the GitHub API (data in the 2nd part of Table~\ref{tab:data-used}) that was performed on a later date than the commit collection in July 2023, 311 repositories (from the repositories initially collected) were not returned by the GitHub API (were indicated as not found in the result), so they were removed from the dataset along with their respective commits in order to guarantee the consistency of the results. Some repositories and some commits were also not returned during the commit change API call in July 2023 (data in the 3rd part of Table~\ref{tab:data-used}): 36 repositories and 11 commits in repositories from the ones initially collected, and these data were therefore also excluded from the dataset. The final dataset size used in the subsequent analysis is shown in the \emph{removed unavailable} column of Table~\ref{tab:dataset-size}.

\begin{table}
\centering
  \caption{Data used in our analysis.}
  \label{tab:data-used}
\scalebox{0.91}{
  \begin{tabular}{lll}
    \toprule
    \textbf{Property} & \textbf{Variable name} & \textbf{Data format} \\
    \hline
    \multicolumn{3}{c}{from GitHub Search API query}\\
    \hline
    repository name &repository.full\_name& string\\
    repository URL &repository.html\_url& string\\
    commit URL &html\_url& string\\
    text of commit &commit.message& string\\
    date and time of commit &commit.committer.date& date\\
    type of repository owner &repository.owner.type&string\\
    \hline
    \multicolumn{3}{c}{from GitHub REST API repository information resource}\\
    \hline
    repository name &full\_name& string\\
    name of organization&organization.login&string\\
    repository size (kB)& size & integer\\
    repository stars & stargazers\_count& integer\\
    number of repository forks& forks\_count & integer\\
    repository main programming language&language&string\\
    \hline
    \multicolumn{3}{c}{from GitHub REST API commit information resource}\\
    \hline
    total changed LOC&stats.total& integer\\
    total added LOC&stats.additions	& integer\\
    total deleted LOC&stats.deletions& integer\\
\hline
\end{tabular}
}
\end{table}

\textbf{Manual verification.} In order to verify that the data we had left after the above preliminary analysis indeed contain references to privacy-relevant laws, we performed a manual verification on a statistically representative sample of the dataset with a 95\% confidence level (\texttt{sample} function of the R programming language): 381 commit messages for GDPR and 320 for CCPA, apart from the case of CPRA and Data Protection Act, where all 69 and 67 commits from the previous step were examined respectively. This process to manually inspect a total of 837 commits was performed by the two authors. Especially for the case of CPRA where we had encountered a large number of commits before the signing of the law (providing an indication that CPRA is used with a different meaning in many cases), we performed a filtering removing commits from specific repositories (e.g. all repositories under the \emph{CPRA-MP} organization). For the case of Data Protection Act commits, a manual filtering was also performed: two commits referring to Data protection Act of countries other than UK were removed, four referring to earlier data protection acts (1998, 2012) (example).\footnote{\url{https://github.com/ckameron99/myNEA/commit/7e78ceb5f3a693455afcb8ec2cb0017985917e4f}} Although the final number of commits for CPRA and Data Protection Act is very small, we kept it in our dataset since CPRA builds upon CCPA and Data Protection Act complements GDPR. For this part of the coding task, there were no disagreements between the two coders: all GDPR and CCPA commits examined were marked as relevant except one case for CCPA where the abbreviation is used differently,\footnote{\url{https://github.com/wosnat/ccpa/commit/81f72876b9908774cfde5dac6f800d1212812385}} even though it is not a clear case as there is no repository description, so it was decided to filter out all commits from the same repository. CPRA and Data Protection Acts commits were filtered with the same output of filtered commits between the two coders. The dataset size, after applying the above filtering, is shown in the \emph{final} column of Table~\ref{tab:dataset-size} with commits coming from a total of 12,391 different repositories. 

\subsection{Pre-processing and analysis tools}

For the analysis on repository and commit level, we removed commits returned more than once (by more than one keyword search and same commits appearing in more than one repositories). However, for some analysis where we compare laws (RQ1, RQ2, RQ3), we kept duplicates to capture all keyword appearances. For the commit comment text analysis (RQ3), we performed the following pre-processing steps commonly encountered in NLP works~\cite{ferreira2024incivility}: 
\begin{itemize}
    \item In order to filter out code, we removed source code blocks marked with backtick characters:\begin{lstlisting}[upquote=true]
``` [...] ```
\end{lstlisting}
    \item We removed HTML code blocks (using HTML tags start and end).
    \item We identified external URLs and removed them.
    \item We removed non-latin characters (the respective commits were however, kept), numbers, punctuation and whitespace.
    \item We converted all words to lowercase.
    \item We removed common English stop words from the commit messages.
    \item We removed the following terms that are frequently encountered in commits, as they were considered noise for our case as being commit terminology and not privacy relevant: \emph{pull}, \emph{request}, \emph{issue}, \emph{fix}, \emph{plugin}, \emph{merge}, \emph{update}, \emph{test}, \emph{branch}, \emph{json}, \emph{release}, \emph{package}, \emph{variable}, \emph{unit}. We relied on common terminology to create the list considering also an analysis on verbs found in commits.\footnote{\url{https://gist.github.com/scmx/411f6fea4ee3832806720d536a7d5d8f}}
\end{itemize}

For our mainly quantitative analysis, we used statistical analysis with descriptive statistics, and text analysis, whereas manual coding was used for the qualitative analysis of RQ4 and the exact process followed is further detailed in the results description of RQ4. Combining both automated and manual analysis allows us to improve the accuracy of our findings. For implementation purposes, libraries of the R programming language were employed (e.g. \texttt{dplyr}, \texttt{wordcloud2}, \texttt{tm}, \texttt{ggplot2}), whereas some of the descriptive statistics, statistical tests and Cohen's kappa agreement calculation were run using IBM SPSS Statistics. As in previous works, we measure the size of a commit by counting the (source code) Lines of Code it affected~\cite{hattori2008nature}. The changed LOC in GitHub considers both the added and the deleted LOC. 

\section{Results}

\subsection{RQ1. Main data privacy laws in commits and time}

\begin{figure}[!t]
\centering
\includegraphics[scale=0.63]{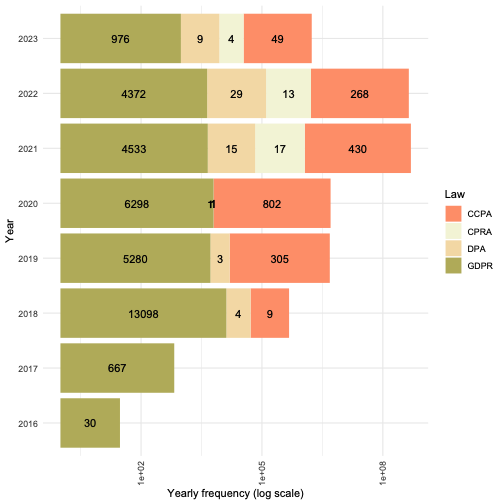}
\caption{Privacy laws appearing over the years counting number of commits.}
\label{fig:laws-per-year} 
\end{figure}

We calculated the commits and total changed LOC for each law per year (number of commits depicted in~\figurename~\ref{fig:laws-per-year} and LOC numbers available in the replication package of the work -- please refer to the Data statement). Although a commit may contain additional changes -- not limited to privacy laws -- the reference to privacy laws provides evidence that a reaction to privacy laws was included in the repository change performed. GDPR is the only law applicable in 2016 and 2017 and has the largest number of commits and changed LOC in 2018 (13,098 or 37.15\% of all commits mentioning GDPR and 40.01\% of all changed LOC for GDPR), which is the year the law came into effect, so it is expected that most repositories decided to perform relevant changes on that year. The changes are also almost equally divided before and after the law effect date for 2018: 6,095 (46.53\%) before and 7,003 (53.47\%) after. For CCPA, most commits are in its effect year 2020 (802 or 43.05\%) but most LOC were changed in 2021 (41.83\%). The effective date of CPRA is very recent in the beginning of 2023, so we might see more relevant commits after the data collection date. For the case of Data Protection Act, most commits and changed LOC appear in 2022 (29 or 47.54\% of commits and 58.30\% of changed LOC) which may be either attributed to the developments in the UK concerning the Data Protection and Digital Information Bill (introduced in the House of Commons in UK on July 18, 2022), or the changes may refer to acts with same naming in other countries that have come into effect in 2021 and 2022. 

\begin{figure}[!t]
\centering
\includegraphics[scale=0.38]{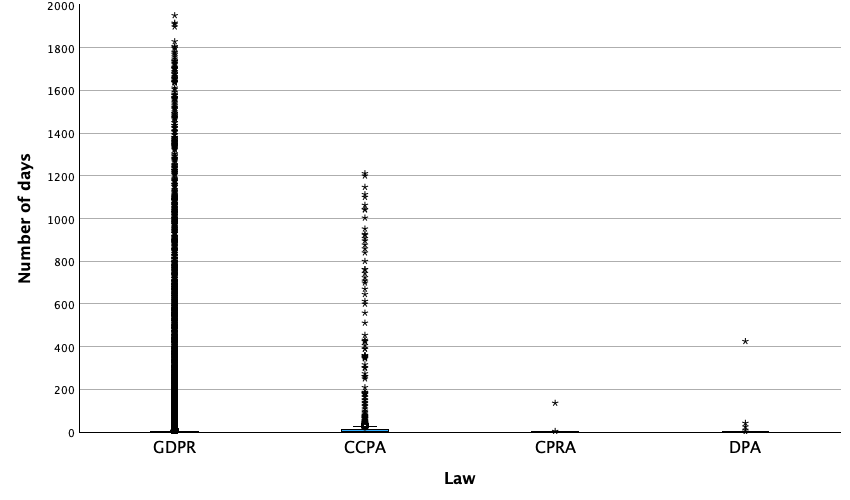}
\caption{Number of days between first and last commit in repository} per law.
\label{fig:days-between-first-and-last-commit} 
\end{figure}

The first and last commit dates across all repositories are shown in Table~\ref{tab:first-last-commits-dates}. The first reference to CPRA is before the introduction of the law, where it appears in the commit message together with GDPR: ``[...]\emph{Add CPRA and GDPR are coming soon}[...]."\footnote{\url{https://github.com/aleecia/CCPABrowserTool/commit/fc795a994d6284633393ca940796ce82fd69f702}} Overall in the dataset there is a small number of commits that refer to more than one laws: 392 commits that mention both CCPA and GDPR which is the most usual case, 10 that mention both CCPA and CPRA, three that mention both DPA and GDPR and one that mentions both CPRA and GDPR. We also examined how long it took for each repository to commit the changes performed by calculating the number of days between relevant commits for each law (using the start and the last dates of relevant commits in each repository). The box plot of~\figurename~\ref{fig:days-between-first-and-last-commit} shows that some repositories have spent more time on making repository adaptations for GDPR, even though the average values are comparable for GDPR and CCPA: for some repositories it took almost 1,900 days (1,951 is the highest value). There is a large deviation in the days among repositories with mean ($M$) days between first and last commit and standard deviation ($SD$) shown in the fifth and sixth column of Table~\ref{tab:first-last-commits-dates}, while for most cases relevant changes were completed within one day (8,743 or 70.56\% of repositories). The above analysis does not consider the time required to perform the actual code change but gives an indication on whether additional changes were necessary in the period after the first relevant commit.

\begin{table}
\centering
  \caption{Dates of first and last commits in dataset and commits duration in repositories per law.}
  \label{tab:first-last-commits-dates}
\scalebox{0.74}{
  \begin{tabular}{ccrrrrrr}
    \toprule
    \textbf{Law} & \textbf{Effect date} & \textbf{First date} & \textbf{Last date (till } & \textbf{$M$} & \textbf{$SD$ in} & \textbf{$M$ total} & \textbf{$SD$ total}\\
    &&&\textbf{April, 2023)} &\textbf{days} & \textbf{days} & \textbf{LOC} & \textbf{LOC}\\
    \hline
    GDPR &May 25, 2018& Apr. 22, 2016 & Mar. 24, 2023&66.90&227.406 & 2,271.58& 48,222.088\\
    CCPA &Jan. 1st, 2020& Jun. 29, 2018& Mar. 19, 2023&76.96&217.066 & 1,084.66& 9,886.801\\
    CPRA &Jan. 1st, 2023& Oct. 29, 2020 & Febr. 24, 2023&17.21&40.276 & 559.77 & 2,462.241\\
    DPA &May 25, 2018& May 24, 2018 & Mar. 1, 2023&23.90&92.462 & 50.67& 59.909\\
\hline
\end{tabular}
}
\end{table}

In terms of LOC affected by the commits, each commit affects on average 2,206.91 LOC ($SD=46,988.575$). As would maybe be expected more lines are added to accommodate changes than deleted: $M=1,720.16$, $SD=44,615.253$ for added LOC and $M=486.74$, $SD=12,885.866$ for deleted LOC. Each commit may also include code changes outside the privacy laws, so the changed LOC numbers may appear higher. We also ran a one-way ANOVA test to check whether the law the committer intends to comply with affects the changes that were performed (using changed LOC), with the hypothesis that GDPR requires more changes. Although the results are not statistically significant, the average changed LOC are more for the case of GDPR, as shown in the last two columns of Table~\ref{tab:first-last-commits-dates}. \\

\noindent\fbox{%
    \parbox{\linewidth}{%
    \textbf{Main findings RQ1:} For GDPR and CCPA, most commits appear on the year the law was put into effect (2018 and 2020 respectively), while in 2018 GDPR changes are almost equally divided before and after the law effect date. Most repositories did not commit changes on days later than the initial commit, but GDPR changes are taking longer (in days) to commit than changes for CCPA, CPRA and DPA. 
    }%
}

\subsection{RQ2. Which type of repositories and which programming languages} are more active

\begin{figure}[!t]
\centering
\includegraphics[scale=0.39]{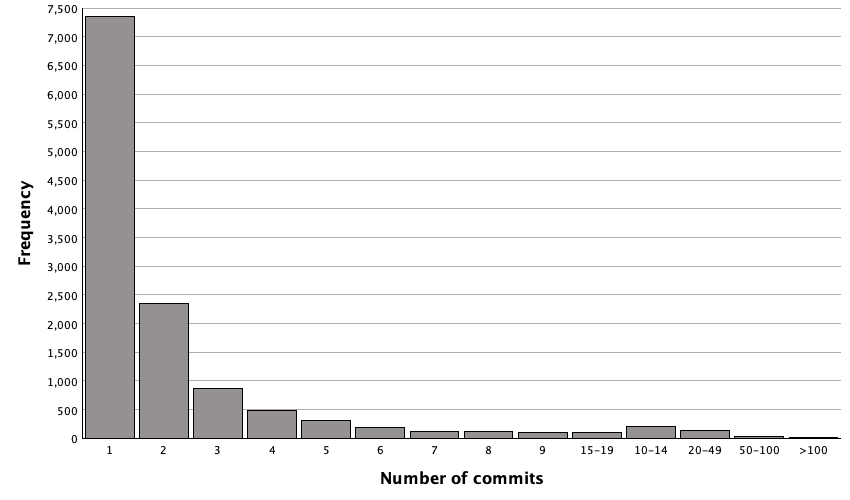}
\caption{Frequency of privacy law relevant commits in repositories.}
\label{fig:number-of-commits-in-repos} 
\end{figure}

In this RQ, we calculated which repositories are more active in privacy law-related commits, considering the number of commits. In terms of commits number, most repositories have only one relevant commit (7,344 or 59.27\% of repositories as in \figurename~\ref{fig:number-of-commits-in-repos}). There are only 54 repositories (0.44\%) with 50 or more relevant commits and only five with more than 300. Among the five most active repositories, one is a repository integrating code from many repositories, one focuses exactly on data compliance (\emph{ministryofjustice/dps-data-compliance}) -- so for those two cases a higher number of commits is expected -- and three are related with popular frameworks: WordPress (\emph{helsinki-systems/wp4nix}), Microsoft 365 documentation (\emph{MicrosoftDocs/microsoft-365-docs}) and prebid programmatic advertising strategy (\emph{8secz-johndpope/PPI}), so the contributors might be more diligent when it comes to privacy compliance. Repositories with only one commit have on average 4,833.67 changed LOC, whereas repositories with more than one commit also have in total more changed LOC: $M=9,061.74$. We had a closer look to those cases of only one commit by manually inspecting a representative sample (366 commits with 95\% confidence level) to verify that they are relevant to privacy law changes and this was confirmed. This process was performed by one of the authors. There are 352 commits (from 227 repositories) in our dataset with zero changed total LOC. With manual inspection on a representative sample of 184 such commits with 95\% confidence level from those repositories, we observed that these cases refer mainly to file deletions and additions, while some are not source code files and for this reason do not have LOC indication, so we kept them since they are also relevant with data privacy laws and useful to answer our RQs (we verified that the commits were relevant to privacy laws --example).\footnote{\url{https://github.com/prmika/ErasmusProject/commit/c8706e85bef44726c7891dbb3344e9e0b3a48028}} This process was also performed by one of the authors.

We examined whether the repository owner type (user or organization) affects the level of activities performed. For this purpose, we run an independent samples t-test using the number of commits as dependent variables and the type of repository owner as independent variable. There is a statistically significant difference showing that commits coming from organization repositories are more than the ones from user repositories ($F=83.357$, $p<0.001$), indicating that organizations perform more changes than individually owned repositories that may be experimental or ones serving educational purposes (descriptive statistics in Table~\ref{tab:commits-repo-type-owner}).

\begin{table}
\centering
  \caption{Number of commits per repository owner type.}
  \label{tab:commits-repo-type-owner}
\scalebox{0.87}{
  \begin{tabular}{crr}
    \toprule
    \textbf{Repository owner type} & \textbf{N} & \textbf{$M$ \# commits}\\
    \hline
    User & 6,351 &2.25\\
    Organization & 6,040 &3.72\\
\hline
\end{tabular}
}
\end{table}

Running ANOVA using as independent variable the main programming language of the repository and as dependent the number of commits and the number of days between the first and the last commit (as used in RQ1), in order to test the hypothesis that the language affects the time required to complete relevant changes, a statistically significant different was found ($F=1.418$, $p=0.002$ for number of commits and $F=4.097$, $p<0.001$ for days between commits). This shows that language constructs might affect the changes required but further investigation is needed to explore other factors, such as framework usage. There are 125 different programming languages indicated as main languages in the repositories we collected with the 10 most frequent programming languages shown in Table~\ref{tab:top-languages}, whereas there is no language indication for 6.6\% of the repositories (819 repositories). The laws that appear in the commits of each programming language, as well as the mean values for the days required to complete relevant commits for data privacy legislation and the commits count are also shown in Table~\ref{tab:top-languages}. Overall, different laws are linked with each programming language.\\

\begin{table}
\centering
  \caption{Top programming languages of repositories.}
  \label{tab:top-languages}
\scalebox{0.87}{
  \begin{tabular}{lrrlrr}
    \toprule
    \textbf{Language} & \textbf{\#} & \textbf{\%} & \textbf{Laws in}&\textbf{$M$}& \textbf{$M$ \# }\\
&\textbf{repos.}&\textbf{repos.}&\textbf{commits}&\textbf{days}&\textbf{commits}\\
    \midrule
    JavaScript & 2,343 &18.9\%&GDPR, CCPA, CPRA,DPA&49.66&2.89\\
    HTML& 1,820 &14.7\%&GDPR, CCPA, 
 CPRA, DPA&52.97&2.49\\
    PHP & 1,737 &14.0\%&GDPR, 
 CCPA&99.60&2.82\\
    Python& 832 &6.7\%&GDPR, CCPA, DPA&96.99&2.91\\
    TypeScript&804 &6.6\%&GDPR, CCPA, CPRA, DPA&67.16&2.66\\
    Java& 635 & 5.1\%&GDPR, CCPA, CPRA&56.40&3.53\\
    CSS& 621 &5.0\%&GDPR, CCPA, CPRA&32.58&2.03\\
    C\#& 375 &3.0\%&GDPR, CCPA&65.03&3.03\\
    Ruby& 301 &2.4\%&GDPR, CCPA&84.14&2.68 \\
    SCSS& 221 &1.8\%&GDPR, CCPA, DPA&56.21&.38\\
\bottomrule
\end{tabular}
}
\end{table}

\noindent\fbox{%
    \parbox{\linewidth}{%
    \textbf{Main findings RQ2:} Most repositories have performed only one commit to accommodate data privacy legislation whereas organization-owned repositories perform more commits. The period between relevant commits is longer for repositories with PHP, Python and Ruby as main programming language.
    }%
}

\subsection{RQ3. Main terms appearing in commits}

\begin{figure*}[h!]
\centering
\begin{subfigure}{0.25\textwidth}
    \includegraphics[width=\textwidth]{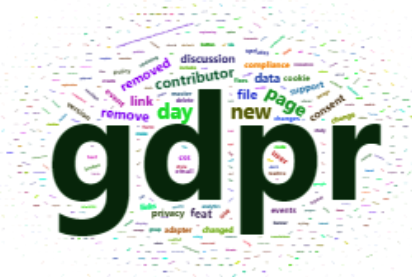}
    \caption{GDPR}
    \label{fig:first}
\end{subfigure}
\hspace{-0.3cm}
\begin{subfigure}{0.25\textwidth}
    \includegraphics[width=\textwidth]{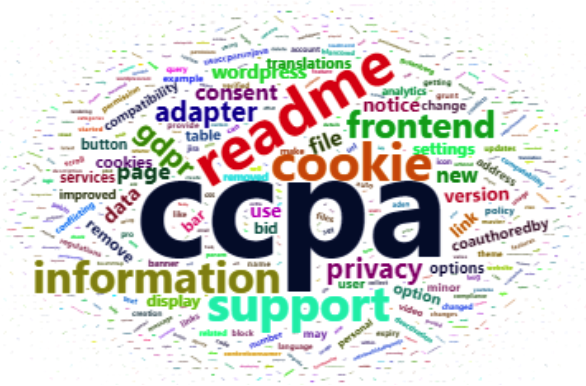}
    \caption{CCPA}
    \label{fig:wordcloud-ccpa}
\end{subfigure}
\hspace{-0.3cm}
\begin{subfigure}{0.25\textwidth}
    \includegraphics[width=\textwidth]{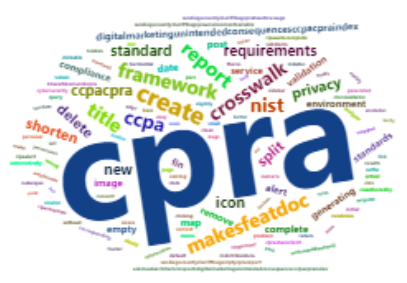}
    \caption{CPRA}
    \label{fig:third}
\end{subfigure}
\begin{subfigure}{0.24\textwidth}
    \includegraphics[width=\textwidth]{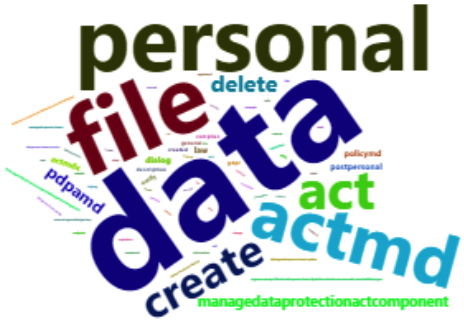}
    \caption{DPA}
    \label{fig:fourth}
\end{subfigure}
\caption{Wordclouds from text of commit messages.}
\label{fig:wordclouds}
\end{figure*}

We examined separately for each law other terms appearing in the commits together with the law name and present the results in the form of wordclouds. We have relied on the pre-processing described in the previous section and used a term frequency (TF) matrix with unigrams for this purpose. All wordclouds are depicted in~\figurename~\ref{fig:wordclouds}. In the wordclouds, we see relevant terms from privacy legislation and required changes: law, delete, privacy, standard, create, new, nist, requirements, analytics, settings, data, consent, remove, compliance, option, notice, change, cookies. More than one law is present in some cases, e.g. GDPR in CCPA relevant commits, CCPA in CPRA relevant commits, so some commits address more than one laws. Although some common words were removed during the pre-processing step, we see some terms on development activities (e.g. contributor).

We also performed a keyword search within the commit message to see if there are any references to the main user rights found in the legislation, as shown with the respective frequencies in Table~\ref{tab:user-rights-in-commits}. The rights are grouped based on the aim of each right, but for searching in the commits we also used different terminology of the rights using a keywords list based as starting point on the list suggested for GDPR in~\cite{vanezi2021complicy}. The main privacy laws share terminology in user rights (first and second column in Table~\ref{tab:user-rights-in-commits}) but we devised variations also for other rights using our prior experience on privacy policies~\cite{DBLP:conf/webist/KailiK22} creating a relevant list, e.g. for \emph{right to opt-out} and \emph{Do Not Sell My Personal Information} for CCPA. The keyword variations for \emph{right to erasure} are listed below and the whole list is available as part of the replication package of the work:
\begin{itemize}
    \item allow deletion, allow (to) delete, right to erasure, right to delete, right to deletion, right of deletion, right of erasure, right to request deletion, right to be forgotten, erase information, request erasure, erase the personal data, erase your personal data, erase their personal data, erase his/her personal data, erase your data, erase their data, erase his/her data, erase any personal data, erase personal data, erase data, right to erase, right to be forgotten, user(s) dalete, user(s) to dalete, data deletion, data delete, delete right, deletion right, deleting right, delete flow, delete request, forget visitor data, forget data
\end{itemize}

We also added the general right of providing consent to data collection that is the case most usually encountered in commit messages (7.35\% of commits). Most other rights appear in a limited number of commits. Only the \emph{right to erasure} (right to request deletion of personal data) is more frequent (1.51\% of all commits, with the word \emph{delete} appearing more times: 1,731 in total), followed by the \emph{right to opt-out} and the \emph{right of access} (right to obtain a copy of personal data).\\ 

\begin{table}
\centering
  \caption{User rights in legislation and frequencies in commits.}
  \label{tab:user-rights-in-commits}
\scalebox{0.9}{
  \begin{tabular}{lcrr}
    \toprule
    \textbf{User right} & \textbf{Legislation} & \textbf{Freq.} \\
    \hline
    General consent (opt-in) & all& 2,706\\
    \hline
    Right to erasure &GDPR, UK DPA&555\\
    Right to delete (or Right to deletion) &CCPA&\\
    \hline
    Right to opt-out &CCPA&153\\
    General consent removal &GDPR, UK DPA& \\
    \hline
    Right of access &GDPR, UK DPA& 84\\
    Right to disclosure & CCPA &\\
    \hline
    Right to rectification &GDPR, UK DPA&46\\
    Right to correct &CPRA& \\
    \hline
    Right to data portability &GDPR, UK DPA, CPRA&20\\
    \hline
    Right to restriction of processing &GDPR& 9\\
    Right to limit (use and disclosure) &CPRA&\\
    Right to restrict processing &UK DPA&\\
    \hline
    Right to object &GDPR, UK DPA&4\\
    \hline
    Right to information &GDPR& 3\\
    Right to know (or Right to notice) &CCPA&\\
    Right to be informed &UK DPA&\\
    \hline
    Right to non-discrimination &CCPA&1\\
    \hline
    Right to avoid automated decision-making &GDPR&0\\
    Right to opt-out of automated decision making&CPRA&\\
\hline
\end{tabular}
}
\end{table}

\noindent\fbox{%
    \parbox{\linewidth}{%
    \textbf{Main findings RQ3:} In the commit comments, there is presence of some privacy related terms but reference to user rights is scarce, with the \emph{right to erasure} being the right encountered more frequently followed by the \emph{right to opt-out} and the \emph{right of access}. User consent provision is the most important aspect commits focus on.
    }%
}

\subsection{RQ4. Which is the main purpose of a commit}

\textbf{Analysis process.} In this RQ, we examined the text of a subset of commit messages and the respective code changes in order to get a better understanding of the purpose of the commit. For the manual analysis of commits text, we selected the 70 GitHub projects with the highest popularity in the dataset using stars as a proxy of similarity as also indicated earlier in the text (Table~\ref{tab:70-most-popular-repos}). The top 70 repositories we focused on have in total 594 unique commits (with 703,862 changed LOC).

The two authors performed the coding task required to answer RQ4. The coding protocol used was as follows. The task of each coder for this part was to read the commit message and the updated files, and classify the commit into one of the following categories: 1) user consent relevant update in source code or text, 2) source code update to cover a user right, 3) privacy policy or FAQ or documentation update, 4) other source code or text update. When present, the pull request linked to the commit was also examined. These categories were created based on the manual verification performed during the dataset creation process when it was observed that some commits refer e.g. to the privacy policy. If the coder encountered however, an additional category she would add it in the list, without consulting the other coder, but in the end the two coders discussed the results. Thus, a bottom-up approach was partially used in the category creation. Each coder would assign each commit to one category as basic category of the commit (even if the commit would also be related to a secondary category, one was provided as main). The independent coding task lasted approximately seven hours for each coder. Both coders are computer scientists with expertise in empirical software engineering research, while one of them has expertise on GDPR.

{\scriptsize
  \begin{longtable}{llrrrrr}
    \toprule
    \textbf{Repository} & \textbf{language} & \textbf{stars}  & \textbf{forks} & \textbf{sum} & \textbf{\#} & \textbf{\#}\\
    &&&& \textbf{LOC} & \textbf{commits} & \textbf{commits}\\
    &&&& \textbf{total} &\textbf{dataset}& \textbf{all}*\\
    \hline
freeCodeCamp/freeCode&TypeScript&	368,467&32,369&396&	3&36,095\\
Camp\\
sindresorhus/awesome	&-&257,402&25,881&1&	1&1,141\\
microsoft/vscode	&TypeScript&147,113&25,886&57,407&	93&124,211\\
microsoft/TypeScript	&TypeScript&91,912	&11,814&69&2&36,238\\
laravel/laravel&PHP	&73,703	&23,744&2&1&7,053\\
gohugoio/hugo	&Go&67,468	&7,219&1,712&8&8,526\\
coder/code-server&TypeScript&	60,731	&5,076&63,721&1&3,831\\
gatsbyjs/gatsby	&JavaScript&54,531&10,538&	1&1&21,674\\
DefinitelyTyped/	&TypeScript&44,257	&29,131&132&1&87,540\\
DefinitelyTyped\\
marktext/marktext	&JavaScript&40,169&	3,063&2&1&1,608\\
discourse/discourse	&Ruby&38,071&8,006&2&	1&55,565\\
iamadamdev/bypass-	&JavaScript&37,917&2,773	&506&7&624\\
paywalls-chrome\\
meilisearch/meilisearch	&Rust&37,002&1,343&	1,243&1&8,796\\
Homebrew/brew	&Ruby&36,231	&8,884&3&1&41,633\\
apache/spark	&Scala&35,949	&27,011&121&1&41,989\\
RocketChat/Rocket.Chat	&TypeScript&35,761&8,616&2,364&	7&26,017\\
getsentry/sentry	&Python&34,403	&3,814&2,140&12&73,159\\
adobe/brackets	&JavaScript&33,439	&7,959&6&1&17,847\\
rapid7/metasploit-&Ruby	&30,410&13,256&292&	2&74,634\\
framework\\
odoo/odoo&JavaScript&	29,409&18,859&4,096&	8&172,835\\
remix-run/remix	&TypeScript&23,724&1,941&9,461&	1&5,884\\
quasarframework/quasar	&JavaScript&23,660&3,211&174&	2&14,349\\
gitlabhq/gitlabhq	&Ruby&23,279&5,734&	114&1&113,301\\
heartcombo/devise	&Ruby&23,211&5,556&24	&1&3,893\\
vuejs/vuepress&JavaScript	&21,611&4,786	&9&3&2,152\\
bazelbuild/bazel&	Java&20,892&3,784&13&	1&40,561\\
forem/forem	&Ruby&20,664&3,736&	1,669&11&13,471\\
goharbor/harbor	&Go&20,234&4,365&92&	1&12,187\\
monicahq/monica	&PHP&19,196&1,910&1,352&	1&809\\
zulip/zulip	&Python&17,961&6,330&250&  	4&60,379\\
WordPress/WordPress	&PHP&17,585&12,292&256&	4&49,941\\
authelia/authelia	&Go&16,581&919	&31&1&6,069\\
JetBrains/intellij-	&-&15,417&4,899&1,855&	13&452,604\\
community\\
chatwoot/chatwoot	&Ruby&15,394&2,231&6&	1&4,338\\
ampproject/amphtml	&JavaScript&14,945	&4,066&2,096&8&22,468\\
hashicorp/packer	&Go&14,450&3,321&776&	1&18,647\\
squidfunk/mkdocs-&HTML	&14,158&2,993&	86&2&6,333\\
material\\
apache/hadoop&	Java&13,584&8,404	&595&6&27,397\\
NodeBB/NodeBB&	JavaScript&13,419&2,709&	422&8&27,507\\
lichess-org/lila&Scala &13,206	&1,942&550&16&61,731\\
Azure/azure-quickstart&Bicep&	13,013&15,649	&69&1&40,527\\
-templates\\
searx/searx	&Python&12,909&1,725&	2&1&4,583\\
networkx/networkx	&Python&12,742&2,929&55&	1&7,769\\
akka/akka	&Scala&12,701&3,606&	34&2&27,076\\
codesandbox/&JavaScript&	12,425&2,190	&2&1&6,401\\
codesandbox-client\\
NixOS/nixpkgs&Nix&	12,270&9,841&	113&7&672,169\\
Automattic/wp-calypso	&JavaScript&12,249&1,998&	23,682&45&69,643\\
segmentio/evergreen&JavaScript	&12,163&858	&887&1&987\\
andkret/Cookbook	&-&11,962&2,234&17	&1&380\\
Chocobozzz/PeerTube&	TypeScript&11,708&1,329&2	&1&13,458\\ 

dotnet/AspNetCore.&C\#&	1,1701&25,554&	4	&3&18,756\\ 
Docs\\
mozilla-mobile/firefox-&Swift	&1,1403&2,720&	221&	1&15,970\\ 
ios\\
overleaf/overleaf	&JavaScript&11,324&1,251&	4&	1&23,222\\ 
matrix-org/synapse	&Python&10,937&2,031&	7,718&	9&23,441\\ 
invertase/react-native-&	JavaScript&10,881&2,168	&1,674&	3&5,815\\ 
firebase\\
airbytehq/airbyte	&Python&10,863	&2,790&152&	1&20,574\\ 
withspectrum/spectrum	&JavaScript&10,744&1,271&	1,254	&6&14,280\\ 
magento/magento2	&PHP&10,735&9,230&	66&	1&146,302\\ 
monkeytypegame/	&TypeScript&10,595&1,570	&22&	2&11,248\\ 
monkeytype\\
knadh/listmonk	&Go&10,185&873	&784&	1&1,426\\ 
ag-grid/ag-grid	&TypeScript&10,118&1,652&	1,587	&5&40,448\\ 
gleitz/howdoi	&Python&10,103&865&	6	&2&982\\ 
abpframework/abp&	C\#&10,026	&3,063&354	&6&36,429\\ 
HabitRPG/habitica	&JavaScript&9,874	&3,707&630	&9&25,439\\ 
MicrosoftDocs/azure-&-	&9,085&19,960	&51,1362&	235&1,283,396\\ 
docs\\
pluja/awesome-privacy	&-&9,066&408&	1	&1&763\\ 
microsoft/sql-server-	&-&8,721	&8,482&0	&2&3,825\\ 
samples\\
languagetool-org/	&Java&8,690	&1,062&5&	2&77,749\\ 
languagetool\\
woocommerce/	&PHP&8,578&10,789&	268&	1&65,102\\ 
woocommerce\\
rolling-scopes/rsschool-	&TypeScript&8,569&186&	5&	1&2,642\\ 
app\\
\cline{1-7}
\multicolumn{7}{r}{\footnotesize{*as of August 2024}}\\
\caption{Top 70 most popular repositories in dataset for manual analysis.}
\label{tab:70-most-popular-repos}
\end{longtable}
}

After the coding task was completed an additional category was added merging categories on cookies/ads/trackers/analytics, as these were indicated by the two coders. They were placed under one category, as they have a similar purpose. This category covers also consent for cookies and cookie preferences. Both coders noticed that some cases were also relevant to security techniques, e.g. encryption, anonymization, but we did not create a separate category due to the small number of such cases (these cases were added in the other source code or text update category). There was an agreement meeting between the two coders, where they discussed cases of disagreement and agreement was reached. Before reaching agreement, Cohen's kappa was $kappa=0.853$ showing a very good agreement between the two coders, as a $kappa$ value higher than 0.81 is considered almost perfect agreement~\cite{viera2005understanding}.

\textbf{Results.} The results are shown in Table~\ref{tab:top-repos-commits-purpose}, along with some examples of commit messages for each category. The vast majority of commits are placed under privacy policy or FAQ or documentation update, whereas some commits accommodate a variety of changes with a mix of changes% including also changes outside data privacy legislation
. We observed that in many of these cases changes performed were in isolated source code files and it was not feasible to link the change with specific content (i.e. specific user rights, law principles) of the law. These changes included, for instance, UI updates, style updates, terminology updates, adding reference to GDPR information.

Some commits refer also to changes so that users of the repository can use adaptations to GDPR, e.g. in \emph{WordPress/WordPress} and \emph{microsoft/vscode} (Visual Studio Code - Open Source) repositories. In the latter repository case, most commits refer to adding GDPR comments or GDPR annotations. In the case of the \emph{MicrosoftDocs/azure-docs} repository, all 235 commits refer to documentation updates, so the number of commits under this category appears higher. Although this is not the typical case, it affects the results of the manual labeling process of this RQ. If we disregard this repository (there are 33 remaining documentation update-relevant commits from 20 of the remaining repositories), the category of source code updates to cover a user right is the most typical.\\

\noindent\fbox{%
    \parbox{\linewidth}{%
    \textbf{Main findings RQ4:} In the top 70 repositories of the dataset in terms of stars count, commits refer to user content updates, user rights updates, policy/FAQ/documentation updates, cookies/ads/trackers/analytics updates, and other updates, with policy/FAQ/documentation updates, other updates and user right updates
    being the most common categories.
    }%
}

\begin{table}[h!]
\centering
  \caption{Main purpose of commits of the 70 most popular repositories in dataset.}
  \label{tab:top-repos-commits-purpose}
  \scalebox{0.83}{
  \begin{tabular}{p{2.2cm}p{1.2cm}p{10.0cm}}
  \toprule
    \textbf{Commit main purpose}&\textbf{commits}&\textbf{Example(s) (repository name: message [SHA])}\\
    \midrule
    \textbf{User consent source} &52& 
    getsentry/sentry:``\emph{feat(gdpr): Prompt new SSO users for marketing consent}" [a80c97e8e2951d3ad637c59178b5c4f7a3dceda1]\\
    \textbf{code/text update}&&JetBrains/intellij-community: ``\emph{GDPR support: improvements in Consents-related UI (layouting \& font)} [70afb5d91d2f713438c799d50613d40839253042]\\
    &&ampproject/amphtml:``\emph{Pass `gdpr\_consent` and `addtl\_consent` to IMA URL}[...]" [3b136eecf60537e96299162830f5e2c263b52fe8]\\
    \textbf{User right source}&67&forem/forem:``\emph{Searchable GDPR Delete Requests Table}[...]" [b155749db06468c4455b6cbf4d9a627236214dc7]\\
    \textbf{code/text update}&&RocketChat/Rocket.Chat:``\emph{[FIX] GDPR action to forget visitor data on request}[...]" [c83a78f5e658186059a130fd31ec2945e5053437]\\
    &&RocketChat/Rocket:``\emph{[FIX] Email sending with GDPR user data}" [1aad2d72443a1d8cd5d2e4a27bc5c4190d770bc8]\\
    &&RocketChat/Rocket.Chat:``\emph{[NEW] GDPR - Right to access and Data Portability}[...]" [fee30ad6f92aa648189757a29b80d1abe78abc40]\\
    \textbf{Policy/FAQ/ document.} &268&freeCodeCamp/freeCodeCamp:``\emph{feat: add gdpr privacy and terms}" [3ad70a7926b1998f0890439da757ea8e637fbeee]\\
    \textbf{update}&&gohugoio/hugo:``\emph{docs: Document the GDPR Privacy Config}" [c71f201fd93287afa7cb7b875bd523c25e48400c]\\
    &&Automattic/wp-calypso:``\emph{Jetpack pricing: add link to GDPR page to footer}[...]" [ccf4f621943d977f75f2bdd533a00372a3efa575]\\
    \textbf{Cookies/ads/ trackers/ analytics} &28&Automattic/wp-calypso:``\emph{Add method to update tracking prefs, limit ccpa checking to US}" [0169ef3fc79d10c9c6a3fe096c659232b446f8de]\\
    \textbf{update}&&Homebrew/brew:``\emph{docs/Analytics: note retention period. Due to [GDPR] https://www.eugdpr.org) Google Analytics have added
[data retention controls]}[...]" [6b3ee9b8fd813c7b1479e629348cb1b096a89819]\\
&&gatsbyjs/gatsby:``\emph{fix(www): fix gdpr for google analytics (\#12098)}" [4fe112cbe708c777d3fc81946f1af752eeff15da]\\
    \textbf{Other source code/text update} &179&getsentry/sentry:``\emph{Update the Jira Integration to comply with the GDPR API changes that disallow using the `name` and `username` keys to search the Jira API}[...]" [643056e6132698be12c3cc420b1cbdb229908efe]\\
    &&apache/hadoop:``\emph{Create Symmetric Key for GDPR}[...]" [46696bd9b0118dc49d4f225d668a7e8cbdd3a6a0]\\
    &&laravel/laravel:``\emph{"Update font delivery (\#5952) Seeing the non compliance of Google Fonts to GDPR I thought to update the CDN.}[...]" [52863d9e4aa41f63592e8c98e5fe717ee7f06a18]\\
    \bottomrule
\end{tabular}
}
\end{table}

\section{Discussion}

\subsection{Findings}

\textbf{Privacy laws indicated.} GDPR is by far the most frequently encountered law. This is expected as it as the law that changed the privacy landscape and started a domino effect towards the update of privacy laws in countries outside EU. CCPA is also more recent than GDPR, but we do not expect relevant commits to reach a comparable number within the next years as we observed that most commits for each law appear close to the law's effect year (RQ1).

\textbf{Number of commits.} Most repositories make reference to privacy legislation in a small number of commits: most repositories performed only one commit and did not devote more than one day in law-relevant commits when we consider the date the commits were performed, as shown in RQ1 and RQ2 results. The number of days between first and last commit is longer for GDPR in some repositories and one reason for this might be that GDPR as the first and main data privacy law caused some uncertainty in how to include some of its provisions in software systems (ambiguity has been identified as a challenge to achieve legal compliance~\cite{li2022towards}). On average, more LOC were added for GDPR that may also be attributed to the fact that subsequent changes (e.g. for CCPA) may have already been covered by GDPR-relevant changes. Comparing these changes with the total number of commits in the most popular repositories of the dataset (RQ4), even repositories with more commit activity than others do not refer to privacy legislation in many commits (Table~\ref{tab:70-most-popular-repos}). This indicates that repositories tend to complete relevant changes in bulk and no need to come back for updates arises. Our dataset covers a large number of programming languages compared to the total number of programming languages on GitHub: in 2022, developers used 500 primary languages according to Octoverse report.\footnote{\url{https://octoverse.github.com/2022/top-programming-languages}} We observed that the average number of commits among programming languages differs but it is not clear whether this is attributed to attributes of the language or whether it is coincidental.

\textbf{User rights indication.} Overall, we found a limited number of user rights in commit messages. Most repositories take actions to ensure users provide their consent on data collection or on the privacy policy overall. The \emph{right to erasure} is the user right mainly found (e.g. indicated with the terms: \emph{delete requests}, \emph{forget visitor data}, \emph{delete flow}, \emph{deletion of users}). Although it is not the only one that requires changes to provide the user the possibility to exercise that right, this result might indicate that it is one of the rights most frequently exercised by the end-users and the developers want to ensure they have a mechanism in place. In the simplest case many user rights can be exercised via e-mail requests but some systems automate the procedure providing dedicated forms. Moreover, obtaining and deleting data are the rights users are mainly aware of, whereas the \emph{right to avoid automated decision-making} is not applicable to most software systems. Other rights found are the \emph{right to opt-out}, \emph{right of access}, and \emph{right to restriction of processing}. The \emph{right to information} is also an important right but is usually handled by adding relevant information in the privacy policy of a system, so keyword based search is less suitable for detecting it in the commit messages. Policy updates are nevertheless, very common according to the results of RQ4.

\textbf{Detailed commit messages.} We observed that developers are not very specific when it comes to the purpose of the changes performed, e.g. many mention only GDPR compliance. This absence of reference may be attributed to the fact that a commit performs a variety of changes to consider the privacy law as a whole without referring to specific parts of the law. It would be useful to have more informative messages to assist other developers on relevant changes, since repositories might also be used for educational purposes~\cite{feliciano2016student}. Previous works have examined characteristics of good commit messages~\cite{tian2022makes}.

\textbf{Cookies.} Cookies are an important area of data collection that requires user consent, although this is not always handled as required. Examining GDPR compliance among 20,218 third-party cookies, it was found that only 12.85\% have a cookie policy where a cookie is mentioned~\cite{fouad2020compliance}. In our dataset, the term \emph{cookie(s)} and its composite terms (e.g. \emph{cookiebanner}, \emph{gdpranalyticscookie}, \emph{cookienotice}, after removing punctuation) appears in 2,384 or 6.48\% of commits, indicating that this is an area where updates for privacy legislation are usually required. The words \emph{policy}, \emph{consent} and \emph{cookie} were identified as main terms that increased in frequency in posts after the introduction of GDPR and CCPA in Reddit discussions on mobile application~\cite{parsons2023understanding}. Cookies were indicated among the categories we identified in the manual analysis in RQ4.

\subsection{Implications for the future}

\textbf{Educational activities.} The focus on specific user rights and the absence of others, as one of the main findings discussed in \emph{user rights indication} in the previous subsection, may show that more information is available about the need of software systems to include those rights (i.e. \emph{right to erasure}, \emph{right to opt-out}, \emph{right of access}, and \emph{right to rectification}) but indicates also the general need for more educational activities towards informing software engineers for data privacy law user rights (in the results of the manual coding of the commits of the top 70 repositories in RQ4, 67 commits cover user rights). Such activities could be integrated in software engineering curricula. A prior work notes the need of legal professionals' education to include a better understanding of the underlying technologies~\cite{corrales2022integrating} but the other direction with software engineers' acquiring deeper legal understanding is also required.

\textbf{Automatic detection of privacy laws.} Detecting automatically whether there is reference to privacy laws in development artifacts (e.g. source code, commit messages, pull requests) can be useful for understanding which aspects of the laws are more present and addressed in software repositories and which are encountered less usually. This can also show the type of features relevant to privacy offered to users by software systems (e.g. how frequent automated decision making actually is). There is an existing line of work that performs similar activities on the text of privacy policies for the detection of the areas covered in the text: for automating the policy text analysis~\cite{harkous2018polisis} or for examining compliance to legislation, such as GDPR~\cite{vanezi2021complicy,DBLP:conf/webist/KailiK22}.

\textbf{Automated privacy recommendations.} Our results show also that tools that assist practitioners in adding privacy safety measures in their source code are necessary. More usable compliance-checking tools were indicated as necessary, when it comes to COPPA, GDPR and CCPA~\cite{alomar2022developers}. Another previous work has analyzed the text of issues in order to recommend good first issues to newcomers with the aim of helping the maintainers of the repository and help newcomers get acquainted with the process~\cite{xiao2022recommending}. Similarly, existing good practices could be suggested by existing contributors to practitioners that attempt to integrate data privacy mechanisms, or other legislation principles, in their code. The area of Privacy as Code concerning also the generation of privacy-friendly code is now at its infancy as indicated by Ferreyra et al.~\cite{ferreyra2024good}.

\textbf{Verifying compliance with privacy laws.} The most active year is the law's effect year but many changes were performed also in subsequent years (\figurename~\ref{fig:laws-per-year}), so developers may need access to appropriate tools to assist them to act sooner. Investigating how Large Language Models overall (e.g. GitHub Copilot, Tabnine) can assist developers towards privacy legislation compliance would be useful~\cite{nguyen2022empirical,dakhel2023github,mastropaolo2023robustness}. The actual compliance with the privacy laws is difficult to detect, and the current work did not verify if the changes performed in the repositories were appropriate for compliance, as the source code was not executed. In a previous work in the area of mobile applications, the actual network traffic of 109 applications that need to comply with CCPA was compared with the data contained in responses to consumers’ access requests~\cite{samarin2023lessons}. It was found that at least 39\% of the applications shared device-specific identifiers and at least 26\% geolocation with third parties but did not disclose this in the request responses. We believe thus, that putting more research efforts towards examining compliance at various levels can assist in guiding end-users to trust or not trust specific software applications. 

\section{Threats to validity}

\textbf{External validity.} It refers to the extent we can generalize our findings and is not expected to affect our work, as we used GitHub, the main source available to perform analysis on our main research question. We did not perform any filtering of the commits based on the total number of repository commits, as performed in previous works~\cite{ferreira2022characterizing}, as we wanted to capture as many commits that refer to data privacy legislation as possible. The low number of commits for CPRA and Data Protection Act in the dataset may be less helpful in drawing conclusions for the specific laws. We made the assumption that commit messages with no country of year indication within this date range are referring to UK Data Protection act. Although some cases may refer to other countries, the commits are referring to privacy legislation and can thus, be considered relevant. In RQ4, the results may differ if a larger sample of commits is examined.

\textbf{Internal validity.} We relied on the commits' content under the assumption that the commits refer to the keywords we were interested in. In order to limit the threat of introducing irrelevant commits, we performed appropriate filtering (i.e. filtering out data of dates earlier than the law introduction date) and manually examined a number of commits (837 commits). We may have missed though false negative cases in the data collection, where the commit message does not include the law name but makes a reference to a relevant issue that describes it. Making a GitHub-wide commits collection and analysis is nevertheless, a tremendous task practically.

\textbf{Construct validity.} It measures the degree to which we measure what we claim. When calculating the time required to perform changes, we were not able to measure the time spent to perform the actual source code changes, as this information is not available. We used instead the commit dates to see how long it took to commit all changes. For CPRA this number of days between first and last commit per law may need to be verified with newer data, since its effect date was close to the data collection time. Changed LOC in each commit may include also other changes not relevant to privacy laws, so LOC numbers reported may be slightly inflated. Both commits number and LOC are reported to limit this threat; the commits numbers correspond to privacy relevant changes, as we did not encounter any other cases during the manual verification step, apart from one case for the CCPA keyword accounting only for 0.12\% of the examined cases. Regarding user rights, we used the terminology from the relevant legislation, so we may have missed references to the rights with different terminology within the message. To limit this threat, we used variations in the terminology that we created relying also on~\cite{vanezi2021complicy} for GDPR terms. 

\textbf{Conclusions validity.} The vast majority of the commits have a reference to GDPR, so there might be a bias in our results towards GDPR. We argue though, that since subsequent privacy legislation relied on GDPR, the conclusions are valid for the current view of the data privacy legislation landscape as a whole.

\section{Conclusions}

In this work, we have examined GitHub commits with reference to privacy laws (GDPR, CCPA, CPRA, UK DPA) since the introduction of GDPR (April 14th, 2016) and till April 2023 (date of main data collection). We have examined how many commits were performed and how many LOC were changed per year, what is the average activity in repositories (RQ1), which characteristics of the repository (i.e. owner type, main programming language) might affect this activity (RQ2), and have also performed text analysis on the commit message (RQ3) including a manual analysis on a limited number of commits (RQ4). As future work, we intend to examine more closely the source code itself. The current work did not verify if the changes performed in the repositories were indeed towards compliance, as the source code was not executed. Exploring how repositories make changes to comply with other recent privacy legislation (e.g. New Zealand's Privacy Act came into effect on December, 1st 2020) would be an interesting work. We finally intend to enhance our findings via a survey with the participation of contributors to the commits we studied and explore automated techniques, such as topic modeling, for a deeper understanding of various topics that appear in commit messages.

\vspace{5mm} %5mm vertical space

\textbf{CRediT authorship contribution statement.} Georgia Kapitsaki: Writing – review \& editing, Writing – original draft, Visualization, Validation, Resources, Methodology, Investigation, Data curation, Conceptualization. Maria Papoutsoglou: Writing – review \& editing, Data curation, Visualization.

\vspace{5mm} %5mm vertical space

\textbf{Data statement.} The dataset used in this work is available for replication purposes online.\footnote{\url{https://doi.org/10.5281/zenodo.15532947}}

\vspace{5mm} %5mm vertical space

\textbf{Funding.} This research did not receive any specific grant from funding agencies in the public, commercial, or not-for-profit sectors.

\bibliography{sample-base}
\bibliographystyle{elsarticle-num}

\end{document}